\documentclass[12pt]{article}

\usepackage[totalwidth=460truept,totalheight=598truept]{geometry}
\usepackage{amsfonts,latexsym,amssymb,amsmath,graphicx,accents,slashed,subfigure}
\usepackage{latexsym}
\usepackage[hidelinks]{hyperref}

\global\arraycolsep=1truept

\begin{document}

\null

\bigskip \phantom{C}

\vskip1truecm

\begin{center}
{\huge \textbf{Fakeons And Microcausality:}}

\vskip.65truecm

{\huge \textbf{Light Cones, Gravitational Waves}}

\vskip.65truecm

{\huge \textbf{And The Hubble Constant}}

\vskip1truecm

\textsl{Damiano Anselmi $^{1,2,3}$ and Antonio Marino $^{1,4}$}

\vskip .1truecm

$^{1}$\textit{Dipartimento di Fisica \textquotedblleft Enrico
Fermi\textquotedblright , Universit\`{a} di Pisa,}

\textit{Largo B. Pontecorvo 3, 56127 Pisa, Italy}

\textit{and }$^{2}$\textit{INFN, Sezione di Pisa,}

\textit{Largo B. Pontecorvo 3, 56127 Pisa, Italy}

$^{3}$damiano.anselmi@unipi.it, $^{4}$antonio.marino7695@gmail.com

\vskip2truecm

\textbf{Abstract}
\end{center}

The concept of fake particle, or \textquotedblleft fakeon\textquotedblright
, allows us to make sense of quantum gravity as an ultraviolet complete
theory, by renouncing causality at very small distances. We investigate
whether the violation of microcausality can be amplified or detected in the
most common settings. We show that it is actually short range for all
practical purposes. Due to our experimental limitations, the violation does
not propagate along the light cones or by means of gravitational waves. In
some cases, the universe even conspires to make the effect disappear. For
example, the positivity of the Hubble constant appears to be responsible for
the direction of time in the early universe.

\vfill\eject

\section{Introduction}

\label{intro}\setcounter{equation}{0}

The idea of fakeon has been put forward in 2017 in refs. \cite%
{LWgrav,fakeons} to ovecome the problem of ghosts in higher-derivative
theories and ensure unitarity. With broader applications, fakeons can be
used even in non-higher-derivative theories, for example when a field has a
positive squared mass, irrespectively of the sign of the pole of its
propagator.

The fakeon is a degree of freedom that can only be virtual. It does not
belong to the physical spectrum. It provides a better understanding of the
Lee-Wick models \cite{leewick,LWQED}, overcoming ambiguities \cite{cutk} and
problems with Lorentz invariance \cite{nakanishi}, and actually leading to
the completion of their formulation \cite{Piva}. Moreover, fakeons allow us
to simplify the proofs of perturbative unitarity in gauge and gravity
theories \cite{scatte}. But their most important application is that they
lead to a consistent, and basically unique, theory of quantum gravity \cite%
{LWgrav,UVQG,absograv}.

One physical consequence due to the fakeons is the violation of causality at
energies larger than their masses. The existence of a relation between
higher-derivatives and violations of microcausality has been known for a
long time. For example, in classical electrodynamics, the runaway solutions
predicted by the Abraham-Lorentz force can be eliminated by renouncing
microcausality \cite{jackson}. An analogous \textquotedblleft
reduction\textquotedblright\ can be implemented in quadratic gravity \cite%
{quadra}. The Lee-Wick models lead to the violation of microcausality
naturally, as realized quite soon \cite{LWQED,lee}. Without
higher-derivatives, a version of electrodynamics that has issues with
causality is the Feynman-Wheeler theory \cite{FW}, which (potentially)
involves the classical analogue of a massless fakeon. However, since a
massless fakeon implies the violations of both microcausality and
macrocausality, Feynman and Wheeler developed a nontrivial \textquotedblleft
absorber-emitter theory\textquotedblright\ to annihilate the effects of the
potential fakeon and recover causality altogether.

In quantum field theory, a satisfactory definition of causality is lacking 
\cite{diagrammar}. Bogoliubov's condition \cite{bogoliubov} is off-shell,
like the Lehmann-Symanzik-Zimmermann requirement that fields commute at
spacelike separated points \cite{LSZ}. At the practical level, the
difficulty is to accurately localize spacetime points working with
relativistic wave packets that describe on-shell particles. In most cases,
we may have to downgrade the violation of causality to an unusual form of
the equations \cite{FLRW}. Nonetheless, if the discrepancies with respect to
the predictions that follow from the ordinary equations can be confirmed or
refuted experimentally, we have a way to make progress, no matter what those
discrepancies are supposed to mean.

The fake degrees of freedom must be projected away, both at the quantum
level and classically. An important feature of the theories that contain
fakeons is that the starting classical action is not the true classical
action \cite{causalityQG}, but an \textquotedblleft
interim\textquotedblright\ local action that provides the basic Feynman
rules. The true classical action is nonlocal and can be obtained after the
quantization, by means of a process of classicization of the interim action.
Forms of violation of microcausality survive the classical limit, due to the
presence of certain \textquotedblleft fakeon averages\textquotedblright\ in
the projected field equations. This fact suggests that it might be possible
to detect the violation experimentally at some point, as a discrepancy with
respect to the ordinary equations, or a fuzziness of the initial conditions,
the time evolution, etc.

In this paper we study the effects of the fakeons on the classical limit of
quantum gravity. Can the violations of microcausality be amplified into
violations of causality? Does Lorentz symmetry spread the effects along the
light cones?\ Can the gravitational waves propagate the effects to long
distances?

Since the theory is Lorentz invariant, the violation occurs for sufficiently
small invariant intervals. Then, it should be possible to propagate it to
arbitrary distances, close enough to the light cones, if we wait for a
sufficient amount of time. This is true in principle, but has no practical
consequences, for reasons related to the poor accuracies of our
measurements. In all realistic situations the violation of microcausality
remains confined within a radius of order $1/m$, where $m$ is the fakeon
mass. To spread it out, we need sources that oscillate with a frequency $%
\omega $ of order $m$, which are not realistic, even if we assume that the
fakeon masses $m$ are relatively small (say, several orders of magnitude
below the Planck mass).

We also find that the impact of fakeons on the gravitational radiation is
negligible, which excludes the possibility of propagating the violation of
microcausality to longer distances by means of gravitational waves. Again,
it would be necessary to generate radiation with very large frequencies.

We conclude that the amplification of the violation of microcausality does
not appear to be around the corner. Actually, nature is for some reason
keeping it confined down to small distances. For example, we show that the
positivity of the Hubble constant conspires to suppress the violation and
give time a direction in the early universe.

The approach to quantum gravity based on the idea of fake particle comes
from perturbative quantum field theory, so the fakeon prescription is
understood at present in momentum space and working perturbatively around
flat space. The problem of fully understanding the fakeon prescription in
coordinate space (beyond the tree level) or around generic backgrounds is
still open at this stage. In some cases, more general backgrounds can be
reached by means of resummations of the results obtained around flat space.
An example of this type is given in section \ref{hubble}.

The paper is organized as follows. In section \ref{capmedia} we study the
fakeon average and discuss its effects close to the light cones. In section %
\ref{caponde} we extend the analysis to the gravitational waves. In section %
\ref{hubble} we study the effects of the Hubble constant on the fakeon
average. Section \ref{conclusions} contains the conclusions and appendix \ref%
{appe} is devoted to the technical aspects of a calculation.

\section{Light cones and microcausality}

\label{capmedia}\setcounter{equation}{0}

In this section we study the classical limit of the fakeon Green function.
In particular, we show that the violation of microcausality does not
propagate along the light cones if the source is slowly varying for time
intervals of order $1/m$, where $m$ is the mass of the fakeon.

Consider the Klein-Gordon equation 
\begin{equation}
\left( 1+\frac{\square }{m^{2}}\right) \phi (x)=J(x),  \label{KG}
\end{equation}%
where $J$ is a real source and $m$ a mass. If $\phi $ is a fakeon, its
solution is \cite{causalityQG} 
\begin{equation}
\phi (x)=\langle J\rangle _{\text{f}}(x)\equiv \int G_{\text{f}}(x-y)J(y)%
\hspace{0.01in}\mathrm{d}^{4}y,  \label{phiF}
\end{equation}%
where $\langle J\rangle _{\text{f}}$ denotes the \textquotedblleft fakeon
average\textquotedblright\ 
\begin{equation}
\langle J\rangle _{\text{f}}=\left. \frac{m^{2}}{\square +m^{2}}\right\vert
_{\text{f}}J\equiv \frac{m^{2}}{2}\left( \left. \frac{1}{\square +m^{2}}%
\right\vert _{\text{ret}}+\left. \frac{1}{\square +m^{2}}\right\vert _{\text{%
adv}}\right) J.  \label{ave}
\end{equation}%
The violation of causality is due to the contributions of the advanced
potentials.

In Fourier transforms, we get 
\begin{equation}
G_{\text{f}}(x)=\mathcal{P}\int \frac{\mathrm{d}^{4}p}{(2\pi )^{4}}\frac{%
-m^{2}\mathrm{e}^{-ip\cdot x}}{p^{2}-m^{2}}=\frac{1}{2}\left[ G_{+i\epsilon
}(x)+G_{-i\epsilon }(x)\right] ,  \label{gf}
\end{equation}%
where $\mathcal{P}$ denotes the principal value and 
\begin{equation*}
G_{\pm i\epsilon }(x)=-\int \frac{\mathrm{d}^{4}p}{(2\pi )^{4}}\frac{m^{2}%
\mathrm{e}^{-ip\cdot x}}{p^{2}-m^{2}\pm i\epsilon }=\frac{m^{3}}{(2\pi )^{2}}%
\frac{K_{1}\left( \pm im\sqrt{x^{2}\mp i\epsilon }\right) }{\sqrt{x^{2}\mp
i\epsilon }}
\end{equation*}%
are the Feynman Green function and its conjugate, $K_{1}$ denoting the
modified Bessel function of the second kind. Each $G_{\pm i\epsilon }(x)$
can be easily evaluated by means of the Wick rotation from the Euclidean
framework. Then formula (\ref{gf}) gives 
\begin{equation}
G_{\text{f}}(x)=\frac{m^{4}}{8\pi ^{2}}\left[ \frac{K_{1}\left( im\sqrt{%
x^{2}-i\epsilon }\right) }{m\sqrt{x^{2}-i\epsilon }}+\frac{K_{1}\left( -im%
\sqrt{x^{2}+i\epsilon }\right) }{m\sqrt{x^{2}+i\epsilon }}\right] .
\label{fakeGreen}
\end{equation}

Observe that $G_{\text{f}}(x-y)$ vanishes for $(x-y)^{2}<0$, so $\phi (x)$
receives contributions only from the past and future light cones 
\begin{equation*}
\mathcal{C}^{\pm }(x)\equiv \left\{ y\;|\;(y-x)^{2}\geqslant 0,\text{ sgn}%
(y^{0}-x^{0})=\pm 1\right\} 
\end{equation*}%
in $x$, equipped with their interiors. Formula (\ref{phiF}) can be written
as 
\begin{equation}
\langle J\rangle _{\text{f}}(x)=\int_{\mathcal{C}^{-}\cup \mathcal{C}^{+}}G_{%
\text{f}}(x-y)J(y)\hspace{0.01in}\mathrm{d}^{4}y  \label{phiF2}
\end{equation}%
and the violation of microcausality is due to the contributions of $\mathcal{%
C}^{+}$.

For $m\sqrt{x^{2}}\gg 1$, we can use the approximation $K_{1}(z)\sim \mathrm{%
e}^{-z}\sqrt{\pi /(2z)}$, which holds for $|z|\gg 1$, arg$(z)\neq \pi $ mod $%
2\pi $. We find the behavior 
\begin{equation}
G_{\text{f}}(x)\sim \frac{m^{5/2}}{4\sqrt{2}\pi ^{3/2}(x^{2})^{3/4}}\cos
\left( {m\sqrt{x^{2}}+\frac{\pi }{4}}\right) ,\quad \quad m\sqrt{x^{2}}\gg 1.
\label{osci}
\end{equation}%
Since $G_{\text{f}}$ rapidly oscillates for $\sqrt{x^{2}}\gg 1/m$, only the
contributions coming from the regions close to the light cones effectively
matter.

In the limit $m\rightarrow \infty $, $G_{\text{f}}(x)$ is localized in the
present, since formula (\ref{ave}) gives 
\begin{equation}
G_{\text{f}}(x)\rightarrow \delta ^{(4)}(x).  \label{delta}
\end{equation}%
The first terms of the expansion around $m=\infty $, which are 
\begin{equation*}
G_{\text{f}}(x)=\left[ 1-\frac{\square }{m^{2}}+\frac{\square ^{2}}{m^{4}}%
+\cdots \right] \delta ^{(4)}(x),
\end{equation*}%
are good indications that, when $m$ is finite, but large, the violation of
causality is short range.

However, in the limit $m\rightarrow 0$, using $K_{1}(z)\sim 1/z$ for $%
z\rightarrow 0$, we get 
\begin{equation}
G_{\text{f}}(x)\underset{|x^{2}|\ll 1/m^{2}}{\longrightarrow }\frac{im^{2}}{%
8\pi ^{2}}\left( \frac{1}{x^{2}+i\epsilon }-\frac{1}{x^{2}-i\epsilon }%
\right) =\frac{m^{2}}{4\pi }\delta (x^{2}),  \label{-206}
\end{equation}%
which shows that close enough to the light cones the violation spreads out
everywhere with no suppression.

The question is: is it correct to talk about a violation of \textit{micro}%
causality? Or does Lorentz invariance raise it to a violation of \textit{%
macro}causality? Events X and Y separated by the same invariant interval $%
(x-y)^{2}$ give equal contributions to the fakeon average. If their distance 
$|\mathbf{x}-\mathbf{y}|$ is small in some inertial frame, it may be
arbitrarily large in other inertial frames. Nevertheless, we are going to
show that the limit $|(x-y)^{2}|\ll 1/m^{2}$ is practically out of reach.

We begin by working in the rest frame of the source. Then we derive a
relativistically improved approximate formula. At that point, we will be
able to consider the effect of shifting to a moving frame.

We first assume that the source is static and pointlike, i.e. 
\begin{equation}
J(x)=J_{0}\delta ^{(3)}(\mathbf{x-\bar{x}}).  \label{2}
\end{equation}%
From the Fourier transform of $G_{\text{f}}$, we easily get the Yukawa
behavior 
\begin{equation*}
\langle J\rangle _{\text{f}}=\frac{m^{2}\mathrm{e}^{-mr}}{4\pi r}J_{0},
\end{equation*}%
where $r=|\mathbf{x-\bar{x}}|$. This result can also be retrieved directly
from (\ref{fakeGreen}) using the tricks explained in the appendix. An
extended static source $J(\mathbf{x})$ gives 
\begin{equation}
\langle J\rangle _{\text{f}}(x)=\frac{m^{2}}{4\pi }\int \mathrm{d}^{3}%
\mathbf{y}\frac{\mathrm{e}^{-m|\mathbf{x}-\mathbf{y}|}}{|\mathbf{x}-\mathbf{y%
}|}J(\mathbf{y}).  \label{3}
\end{equation}%
The sensitivity of the average to the source is exponentially damped by the
distance from the observer.

Next, a pointlike source oscillating with frequency $\omega $, 
\begin{equation}
J(x^{0},\mathbf{x})=J_{0}\mathrm{e}^{-i\omega x^{0}}\delta ^{(3)}(\mathbf{x-%
\bar{x}}),  \label{furia}
\end{equation}%
gives 
\begin{equation}
\langle J\rangle _{\text{f}}(x)=\frac{J_{0}m^{2}\mathrm{e}^{-i\omega x^{0}}}{%
4\pi r}\left\{ 
\begin{tabular}{ll}
$\mathrm{e}^{-r\sqrt{m^{2}-\omega ^{2}}}$ & \qquad for $\omega <m,$ \\ 
$\cos \left( r\sqrt{\omega ^{2}-m^{2}}\right) $ & \qquad for $\omega >m.$%
\end{tabular}%
\ \right.  \label{omega}
\end{equation}%
Note that the fakeon prescription is needed only for $\omega >m$. For $%
\omega \gg m$ the approximate behavior 
\begin{equation*}
\langle J\rangle _{\text{f}}(x)\sim \frac{J_{0}m^{2}}{8\pi r}\left[ \mathrm{e%
}^{-i\omega (x^{0}-r)}+\mathrm{e}^{-i\omega (x^{0}+r)}\right] \ 
\end{equation*}%
shows that the violation of microcausality does propagate along the light
cones.

To treat the most general case, let us introduce the Fourier transform 
\begin{equation}
J(x^{0},\mathbf{x})=\int \frac{\mathrm{d}\omega }{2\pi }\mathrm{e}^{-i\omega
x^{0}}\tilde{J}(\omega ,\mathbf{x})  \label{trasfa}
\end{equation}%
with respect to time. From (\ref{gf}) and (\ref{omega}), we find 
\begin{eqnarray}
\langle J\rangle _{\text{f}}(x) &=&\int_{-m}^{m}\frac{\mathrm{d}\omega }{%
2\pi }\mathrm{e}^{-i\omega x^{0}}\int \mathrm{d}^{3}\mathbf{y}\frac{m^{2}%
\mathrm{e}^{-\sqrt{m^{2}-\omega ^{2}}|\mathbf{x}-\mathbf{y}|}}{4\pi |\mathbf{%
x}-\mathbf{y}|}\tilde{J}(\omega ,\mathbf{y})  \notag \\
&&+\int_{|\omega |\geqslant m}\frac{\mathrm{d}\omega }{2\pi }\mathrm{e}%
^{-i\omega x^{0}}\int \mathrm{d}^{3}\mathbf{y}\frac{m^{2}\cos (\sqrt{\omega
^{2}-m^{2}}|\mathbf{x}-\mathbf{y}|)}{4\pi |\mathbf{x}-\mathbf{y}|}\tilde{J}%
(\omega ,\mathbf{y}).  \label{comple}
\end{eqnarray}%
This result shows that all the frequencies $\omega <m$ are exponentially
damped by the distance between the observer and the source. Instead, no
frequencies $|\omega |\geqslant m$ are damped.

If the source $J(y^{0},\mathbf{y})$ is slowly varying in an amount of time
comparable to $1/m$, i.e. 
\begin{equation}
\tilde{J}(\omega ,\mathbf{y})=0\text{ for }\omega >\bar{\omega}\text{, for
some }\bar{\omega}\ll m,  \label{appro}
\end{equation}%
the violation of causality can be estimated by comparing the exact solution 
\begin{equation}
\phi (x)=\langle J\rangle _{\text{f}}(x)=\int_{-\bar{\omega}}^{\bar{\omega}}%
\frac{\mathrm{d}\omega }{2\pi }\mathrm{e}^{-i\omega x^{0}}\int \mathrm{d}^{3}%
\mathbf{y}\frac{m^{2}\mathrm{e}^{-\sqrt{m^{2}-\omega ^{2}}|\mathbf{x}-%
\mathbf{y}|}}{4\pi |\mathbf{x}-\mathbf{y}|}\tilde{J}(\omega ,\mathbf{y})
\label{exa}
\end{equation}%
to the causal formula 
\begin{equation}
\langle J\rangle _{\text{f}}(x)\sim \int \mathrm{d}^{3}\mathbf{y}\frac{%
m^{2}J(x^{0}-|\mathbf{x}-\mathbf{y}|,\mathbf{y})}{4\pi |\mathbf{x}-\mathbf{y}%
|}\exp \left( -\sqrt{m^{2}-\bar{\omega}^{2}}|\mathbf{x}-\mathbf{y}|\right)
\equiv \tilde{\phi}_{\text{c}}(x),  \label{appr}
\end{equation}
which involves the source $J$ only at an earlier, retarded time.

To study the accuracy of the causal approximation, let us focus on a
pointlike source 
\begin{equation}
J(x^{0},\mathbf{x})=j(x^{0})\delta ^{(3)}(\mathbf{x-\bar{x}}),  \label{pola}
\end{equation}%
such that $j(x^{0})$ is an L$^{2}$ function and its Fourier transform $%
\tilde{j}(\omega )$ vanishes for $\omega >\bar{\omega}$. Then, writing the
exact solution (\ref{exa}) as $\phi \equiv \tilde{\phi}_{\text{c}}+\Delta 
\tilde{\phi}$, we have 
\begin{equation}
\mathcal{E}\equiv \frac{||\Delta \tilde{\phi}||}{||\tilde{\phi}_{\text{c}}||}%
\leqslant \frac{\bar{\omega}\sqrt{m+\bar{m}}}{\sqrt{2}\bar{m}^{3/2}}\sim 
\frac{\bar{\omega}}{m},  \label{bound}
\end{equation}%
where $\bar{m}\equiv \sqrt{m^{2}-\bar{\omega}^{2}}$ and $||\cdots ||$
denotes the L$^{2}$ norm. This bound gives us a way to estimate the
precision of the approximation at the global level. Later we discuss the
accuracy of the approximation as a function of the distance $r$ between the
source and the observer, to show that the correction $\Delta \tilde{\phi}$,
which encodes the violation of microcausality, is negligible for all
practical purposes.

\subsubsection*{Improved approximation}

The Green function (\ref{fakeGreen}) is invariant under time reversal T.
Instead, the Green function implied by the causal approximation (\ref{appr})
is not. A better approximation is given by the T-symmetric formula 
\begin{equation}
\langle J\rangle _{\text{f}}(x)\sim \int \mathrm{d}^{3}\mathbf{y}\frac{m^{2}%
\left[ J(x^{0}-|\mathbf{x}-\mathbf{y}|,\mathbf{y})+J(x^{0}+|\mathbf{x}-%
\mathbf{y}|,\mathbf{y})\right] }{8\pi |\mathbf{x}-\mathbf{y}|}\mathrm{e}^{-%
\sqrt{m^{2}-\bar{\omega}^{2}}|\mathbf{x}-\mathbf{y}|}\equiv \tilde{\phi}_{%
\text{T}}(x).  \label{appr2}
\end{equation}%
If we take the source (\ref{pola}), with the assumption (\ref{appro}), it is
easy to check that 
\begin{equation}
\frac{||\tilde{\phi}_{\text{T}}-\tilde{\phi}_{\text{c}}||}{||\tilde{\phi}_{%
\text{c}}||}\leqslant \frac{\bar{\omega}}{\sqrt{2}\bar{m}}\sim \frac{\bar{%
\omega}}{\sqrt{2}m}.  \label{bound3}
\end{equation}%
This means that, globally, the T-symmetric approximation is as good as the
causal one. Neither (\ref{exa}), nor (\ref{appr}), nor (\ref{appr2}) are
Lorentz invariant, due to the assumption (\ref{appro}). However, (\ref{appr2}%
) can be used to switch easily to a moving inertial frame (see below).

\subsubsection*{Properties}

We list a few properties of the causal approximation (\ref{appr}), the
T-symmetric approximation (\ref{appr2}) and the exact solution (\ref{exa}).

1) Formula (\ref{appr2}) has the right behavior for $m\rightarrow 0$, 
\begin{equation}
\lim_{m\rightarrow 0}\frac{\tilde{\phi}_{\text{T}}(x)}{m^{2}}=\frac{1}{8\pi }%
\int \frac{\mathrm{d}^{3}\mathbf{y}}{|\mathbf{x}-\mathbf{y}|}\left[ J(x^{0}-|%
\mathbf{x}-\mathbf{y}|,\mathbf{y})+J(x^{0}+|\mathbf{x}-\mathbf{y}|,\mathbf{y}%
)\right] ,  \label{m0}
\end{equation}%
in agreement with (\ref{-206}), even if, strictly speaking, $m=0$ is not
compatible with the condition of slow variation for time intervals of order $%
1/m$. Formula (\ref{appr}) does not share this property.

2) When $m$ is large, we correctly get 
\begin{equation*}
\lim_{m\rightarrow \infty }\tilde{\phi}_{\text{c}}(x)=\lim_{m\rightarrow
\infty }\tilde{\phi}_{\text{T}}(x)=J(x)
\end{equation*}%
in both cases.

3) If $J(y^{0},\mathbf{y})$ has compact support in $\mathbf{y}$ [$J(y^{0},%
\mathbf{y})=0$ for $|\mathbf{y}|>\rho $, where $\rho $ is some finite
radius] and the Fourier transform $\tilde{J}(\omega ,\mathbf{y})$ vanishes
for $\omega >\bar{\omega}$, $\bar{\omega}\leqslant m$, and is bounded [$|%
\tilde{J}(\omega ,\mathbf{y})|\leqslant K$ for every $\omega $ and $\mathbf{y%
}$], then $\phi (x)$, $\tilde{\phi}_{\text{c}}(x)$ and $\tilde{\phi}_{\text{T%
}}(x)$ tend to zero exponentially at spatial infinity. Indeed, let $|\mathbf{%
x}|>r$ for a $r>\rho $. Then, $|\mathbf{x}|>r>\rho >\mathbf{|y}|$ implies $|%
\mathbf{x}-\mathbf{y}|>|\mathbf{x}|-|\mathbf{y}|>r-\rho $, so (\ref{exa}), (%
\ref{appr}) and (\ref{appr2}) give 
\begin{equation}
|\phi |,|\tilde{\phi}_{\text{T}}|,|\tilde{\phi}_{\text{c}}|<\frac{m^{2}K\bar{%
\omega}\rho ^{3}\mathrm{e}^{-\bar{m}(r-\rho )}}{3\pi (r-\rho )}.
\label{bond}
\end{equation}

\subsubsection*{Data and validity of the approximations}

Now we analyse the experimental data to justify the approximations. The
theory of quantum gravity that emerges from the fakeon idea predicts a
spin-2 fakeon $\chi _{\mu \nu }$ of mass $m_{\chi }$ and a potential scalar
fakeon $\phi $ of mass $m_{\phi }$ (see the next section for details). The
masses $m_{\chi }$ and $m_{\phi }$ are free parameters. At present, we do
not have strong bounds on their values. For example, the gravitational
potential of a point-like mass $M$ is 
\begin{equation*}
V(r)=-\frac{GM}{r}\left( 1-\frac{4}{3}\mathrm{e}^{-m_{\chi }r}+\frac{1}{3}%
\mathrm{e}^{-m_{\phi }r}\right) .
\end{equation*}%
Since Newton's law has been verified down to $10^{-2}$cm \cite{eotvos}, we
infer that 
\begin{equation}
m_{\chi },m_{\phi }\gtrsim \frac{10^{2}}{\text{cm}}\sim 10^{-3}\text{eV.}
\label{pote}
\end{equation}%
The fakeon masses could have any values larger than this one. If $m_{\chi
},m_{\phi }$ are smaller than the Planck mass, the quantum gravity theory of
ref. \cite{LWgrav} is perturbative and may allow us to test new physics at
relatively low energies. In what follows, we take $m_{\chi },m_{\phi }\sim
10^{12}$GeV, which means $1/m_{\chi },1/m_{\phi }\sim 10^{-36}$s, as\
reference values. We also compare the results we obtain with those implied
by other values of $m_{\chi },m_{\phi }$.

The shortest time interval that has been measured directly comes from laser
pulses and is about $10^{-17}$s \cite{laserpulses}. We take this value as
the reference amount of time $1/\bar{\omega}$ when we require that the
source is slowly varying at the practical level. If $m_{\chi },m_{\phi }\sim
10^{12}$GeV, the overall accuracy (\ref{bound}) of the approximation (\ref%
{appr}) is 
\begin{equation}
\mathcal{E}=\frac{||\Delta \tilde{\phi}||}{||\tilde{\phi}_{\text{c}}||}\sim 
\frac{\bar{\omega}}{m}\sim 10^{-19}.  \label{e1}
\end{equation}%
We need $m_{\chi },m_{\phi }\sim 70$eV to have $||\Delta \tilde{\phi}||\sim
||\tilde{\phi}_{\text{c}}||$ and make the violation of causality important.

Even in that case, however, we have to fight against the damping exponential
factors $\sim \mathrm{e}^{-mr}$ of formulas (\ref{exa}), (\ref{appr}), (\ref%
{appr2}) and (\ref{bond}), which depress the outcome. Due to them, the
region where the magnitude of $\phi $ is not depressed is a ball of radius $%
1/m_{\chi },1/m_{\phi }\sim 10^{-26}$cm. However, such distances are
unreachable to us. Indeed, the shortest distance ever measured, provided by
LIGO's technology, is around $10^{-17}$cm \cite{LIGO}, where the damping
factor is $\sim \mathrm{e}^{-10^{9}}$.

If we want a damping factor of order unity at the shortest distance ever
measured, we need $m_{\chi },m_{\phi }\sim 2$TeV. However, in that case $%
\mathcal{E}\sim 10^{-11}$ is still too small: the causal formula (\ref{appr}%
) continues to be good enough and the violation of microcausality remains
undetectable.

If the Higgs boson is a fakeon \cite{fakeH}, the violation of microcausality
extends to an amount of time equal to $1/m_{H}\sim 10^{-26}$s, which is
still too short. In that case, $\mathcal{E}\sim 5\cdot 10^{-10}$ and if we
want a damping factor of order one we need to go to distances $r\sim
1/m_{H}\sim 10^{-16}$cm.

Note that, to some extent, it makes sense to assume that the source is
slowly varying in time and not assume that it is weakly varying in space,
since, as seen, the measurements of space distances are much more precise
than those of time intervals. Clearly, the assumption that the source is
slowly varying depends on the reference frame and so implies a
\textquotedblleft spontaneous\textquotedblright\ breaking of Lorentz
symmetry. Can the violation of microcausality be enhanced by switching to a
Lorentz frame that moves at a great speed with respect to the source? In
principle yes, but in practice no.

To see this, we use the approximation (\ref{appr2}). Let $s$ denote the
frame where the assumption (\ref{appro}) holds. If we go to a Lorentz frame $%
s^{\prime }$ that moves at a speed $\beta >0$ relatively to the source $J$
in $s$, the frequencies of $J$ are enhanced by factors 
\begin{equation}
\sqrt{\frac{1-\beta }{1+\beta }},\qquad \sqrt{\frac{1+\beta }{1-\beta }}.
\label{fac}
\end{equation}

If we switch to an inertial frame with $\beta =1-\varepsilon $, $\varepsilon
\ll 1$, the larger factor (\ref{fac}) is $\sim \sqrt{2/\varepsilon }$. If
the maximal frequency of the source $J$ is $\bar{\omega}^{\prime }$, as seen
from $s^{\prime }$, the maximal frequency seen from $s$ is $\bar{\omega}\sim 
\bar{\omega}^{\prime }\sqrt{2/\varepsilon }$. Then, if we assume $m_{\chi
},m_{\phi }\sim 10^{12}$GeV and take $1/\bar{\omega}^{\prime }$ around the
shortest amount of time that has been measured directly so far, which means $%
\bar{\omega}^{\prime }\sim 10^{17}$Hz, we need $\varepsilon \sim 10^{-38}$
to make the right-hand sides of (\ref{bound}) and (\ref{bound3}) of order
one, which is out of reach. Similar conclusions hold with the other values
of $m_{\chi },m_{\phi }$ considered above and for the transverse Doppler
effect.

The second question is: can we reduce the effects of the damping factor $%
\mathrm{e}^{-mr}$? The answer is, again, no, because measurements of space
distances mix with measurements of time intervals, which are much less
precise. Consider the contraction of lengths in special relativity. The
endpoints of a bar of length $l$ at rest in $s$ must be recorded
simultaneously in $s^{\prime }$. However, if the highest precision of a
measurement of time in $s^{\prime }$ is $\Delta t^{\prime }\sim $ $10^{-17}$%
s, then the length 
\begin{equation*}
l^{\prime }=\frac{l}{\gamma }\pm |\beta |\Delta t^{\prime }
\end{equation*}%
of the bar in $s^{\prime }$ has an error that cannot be reduced below $\sim
10^{-7}$cm for $|\beta |\rightarrow 1$. On thop of this, the factor $\gamma $
is practically one for all known macroscopic objects of our galaxy (and far
beyond), which have maximum velocities of order 1000km/s with respect to the
CMB rest frame. Large values of $\gamma $ are hard to reach even for
elementary particles, because we would need to accelerate them to about $%
10^{12}$GeV.

In conclusion, there is no obvious way of reducing the shortest time
intervals or space distances we can measure by switching to different
Lorentz frames. When the source is slowly varying in time the fakeon average
is short range. If, in addition, $J$ has compact support in space, $\langle
J\rangle _{\text{f}}$ is exponentially decreasing at spatial distances.
Under these circumstances, the violation of microcausality is a sort of
fuzziness of the source and its evolution.

A last resort to amplify the violations is to get help from radiation, i.e.
signals that do propagate along the light cones. The electromagnetic
radiation is not very helpful here, since only gravity is sensitive to the
averaged source $\langle J\rangle _{\text{f}}$, while the other interactions
are sensitive to the unaveraged source $J$. In the next section we show that
the gravitational radiation predicted by the classicization of quantum
gravity coincides with the one predicted by the Einstein theory.

\section{Gravitational waves}

\label{caponde}

\setcounter{equation}{0}

Now we study the effects on the gravitational waves. We assume that the
sources are smooth, have compact support in space and are slowly varying for
time intervals of order $1/m$, where $m$ is the fakeon mass.

We recall that the quantum gravity theory of ref. \cite{LWgrav} describes a
triplet made of the graviton, a spin-2 fakeon $\chi _{\mu \nu }$ of mass $%
m_{\chi }$ and a scalar field $\phi $ of mass $m_{\phi }$, which can be fake
or physical. This leads to two physically inequivalent theories, the GFF
(graviton-fakeon-fakeon) theory and the GSF (graviton-scalar-fakeon) theory.
In this section we show that the gravitational waves predicted by both
versions coincide with the ones of Einstein gravity for all practical
purposes. For definiteness, we first work in the GFF theory and then extend
the results to the GSF\ theory.

Neglecting the cosmological constant, the unprojected field equations that
follow from the interim classical action are \cite{causalityQG}

\begin{equation}
\left( 1+\frac{\nabla ^{2}}{m_{\chi }^{2}}\right) G_{\mu \nu }+\frac{r_{\phi
\chi }}{3}\left( \nabla _{\mu }\nabla _{\nu }-g_{\mu \nu }\nabla ^{2}\right)
G=\kappa ^{2}T_{\mu \nu },  \label{eom}
\end{equation}%
where $G_{\mu \nu }$ is the Einstein tensor, $r_{\phi \chi }=(m_{\phi
}^{2}-m_{\chi }^{2})/(m_{\phi }^{2}m_{\chi }^{2})$, $\kappa =\sqrt{8\pi G}$
and 
\begin{equation}
\kappa ^{2}T_{\mu \nu }\equiv \kappa ^{2}T_{\mathfrak{m}\mu \nu }+\frac{1}{%
2m_{\chi }^{2}}g_{\mu \nu }R^{\rho \sigma }R_{\rho \sigma }-\frac{2}{m_{\chi
}^{2}}R_{\mu \rho \nu \sigma }R^{\rho \sigma }+\frac{2m_{\phi }^{2}+m_{\chi
}^{2}}{3m_{\chi }^{2}m_{\phi }^{2}}R\left( R_{\mu \nu }-\frac{1}{4}g_{\mu
\nu }R\right) ,  \label{tmn}
\end{equation}%
$T_{\mathfrak{m}\mu \nu }$ denoting the matter energy-momentum tensor.

Once the fakeons $\phi $ and $\chi _{\mu \nu }$ are projected away, the
field equations for the graviton field $h_{\mu \nu }$, defined as the
fluctuation around flat space by means of the relation $g_{\mu \nu }=\eta
_{\mu \nu }+2\kappa h_{\mu \nu }$, have the form: 
\begin{equation}
G_{\mu \nu }=\kappa ^{2}T_{\mu \nu }^{\text{GFF}},  \label{21}
\end{equation}%
where $T_{\mu \nu }^{\text{GFF}}$ contains the corrections to the Einstein
equations. Since the right-hand side depends on the metric, the equations
have to be treated perturbatively in $\kappa $. We expand the Einstein
tensor as 
\begin{equation*}
G_{\mu \nu }=\kappa \left[ -\square \tilde{h}_{\mu \nu }+\partial _{\mu
}\partial ^{\rho }\tilde{h}_{\rho \nu }+\partial _{\nu }\partial ^{\rho }%
\tilde{h}_{\rho \mu }-\eta _{\mu \nu }\partial ^{\rho }\partial ^{\sigma }%
\tilde{h}_{\rho \sigma }\right] +\kappa ^{2}J_{\mu \nu },
\end{equation*}%
where 
\begin{equation*}
\tilde{h}_{\mu \nu }=h_{\mu \nu }-\frac{1}{2}\eta _{\mu \nu }h,\qquad
h=h_{\mu \nu }\eta ^{\mu \nu },
\end{equation*}%
$\square =\partial ^{2}$ and $J_{\mu \nu }$ is at least quadratic in $%
h_{\alpha \beta }$. Using the definition of $T_{\mu \nu }^{\text{GFF}}$
given in ref. \cite{causalityQG}, equation (\ref{21}) can be recast in the
perturbative form 
\begin{equation}
-\square \tilde{h}_{\mu \nu }+\partial _{\mu }\partial ^{\rho }\tilde{h}%
_{\rho \nu }+\partial _{\nu }\partial ^{\rho }\tilde{h}_{\rho \mu }-\eta
_{\mu \nu }\partial ^{\rho }\partial ^{\sigma }\tilde{h}_{\rho \sigma
}=\kappa \;\left\langle T_{\mu \nu }-U_{\mu \nu }+\frac{r_{\phi \chi }}{3}%
\left( \eta _{\mu \nu }\partial ^{2}-\partial _{\mu }\partial _{\nu }\right)
\left\langle T-U\right\rangle _{\phi }\right\rangle _{\chi },  \label{22}
\end{equation}%
where $\kappa ^{2}U_{\mu \nu }$ is the left-hand side of (\ref{eom}) minus
its linear part, $T=\eta ^{\mu \nu }T_{\mu \nu }$, $U=\eta ^{\mu \nu }U_{\mu
\nu }$ and the fakeon average $\langle \cdots \rangle _{\text{f}}$ is
defined as in formula (\ref{ave}), the masses being $m^{2}=m_{\chi }^{2}$ or 
$m^{2}=m_{\phi }^{2}$, depending on the case.

It is convenient to impose the gauge-fixing condition $\partial ^{\mu }%
\tilde{h}_{\mu \nu }=\partial _{\nu }V$, where $V$ \ is a function to be
determined, because then the equations (\ref{22}) take the form 
\begin{equation}
\square \left( \tilde{h}_{\mu \nu }-\eta _{\mu \nu }V\right) =-\kappa
\langle T_{\mu \nu }-U_{\mu \nu }\rangle _{\chi }+\left( \partial _{\mu
}\partial _{\nu }-\eta _{\mu \nu }\square \right) \left[ 2V+\frac{\kappa
r_{\phi \chi }}{3}\left\langle \langle T-U\rangle _{\phi }\right\rangle
_{\chi }\right]  \label{23}
\end{equation}
and if we choose 
\begin{equation*}
V=-\frac{\kappa r_{\phi \chi }}{6}\left\langle \langle T-U\rangle _{\phi
}\right\rangle _{\chi },
\end{equation*}
they reduce to 
\begin{equation}
\square \left( \tilde{h}_{\mu \nu }+\frac{\kappa \eta _{\mu \nu }r_{\phi
\chi }}{6}\left\langle \langle T-U\rangle _{\phi }\right\rangle _{\chi
}\right) =-\kappa \langle T_{\mu \nu }-U_{\mu \nu }\rangle _{\chi }.
\label{cc}
\end{equation}
\ 

Applying the definition (\ref{ave}) in momentum space, it is easy to prove
the identity 
\begin{equation*}
r_{\phi \chi }\left\langle \langle T-U\rangle _{\phi }\right\rangle _{\chi }=%
\frac{1}{m_{\chi }^{2}}\langle T-U\rangle _{\chi }-\frac{1}{m_{\phi }^{2}}%
\langle T-U\rangle _{\phi }.
\end{equation*}
Inserting this result into (\ref{cc}), we obtain the equation 
\begin{equation}
\square \left( \tilde{h}_{\mu \nu }+\frac{\kappa \eta _{\mu \nu }}{6m_{\chi
}^{2}}\langle T-U\rangle _{\chi }-\frac{\kappa \eta _{\mu \nu }}{6m_{\phi
}^{2}}\langle T-U\rangle _{\phi }\right) =-\kappa \langle T_{\mu \nu
}-U_{\mu \nu }\rangle _{\chi }.  \label{25o}
\end{equation}

We concentrate on the first order in $\kappa $, where $U_{\mu \nu }=J_{\mu
\nu }=0$. Since (\ref{25o}) implies $\tilde{h}_{\mu \nu }=\mathcal{O}(\kappa
)$, we have $R^{2}\sim R_{\mu \nu }R^{\mu \nu }\sim \kappa ^{4}$. Then,
formula (\ref{tmn}) implies $T_{\mu \nu }=T_{\mathfrak{m}\mu \nu }+\mathcal{O%
}(\kappa ^{2})$. At the end, the approximation leads to the equation 
\begin{equation}
\square \left( \tilde{h}_{\mu \nu }+\frac{\kappa \eta _{\mu \nu }}{6m_{\chi
}^{2}}\langle T_{\mathfrak{m}}\rangle _{\chi }-\frac{\kappa \eta _{\mu \nu }%
}{6m_{\phi }^{2}}\langle T_{\mathfrak{m}}\rangle _{\phi }\right) =-\kappa
\langle T_{\mathfrak{m}\mu \nu }\rangle _{\chi }.  \label{25}
\end{equation}

It is convenient to decompose 
\begin{equation*}
\tilde{h}_{\mu \nu }=\tilde{h}_{\mu \nu }^{E}+\tilde{h}_{\mu \nu }^{\text{f}}
\end{equation*}%
as the sum of the solution 
\begin{equation*}
\tilde{h}_{\mu \nu }^{E}(x^{0},\mathbf{x})=-\frac{\kappa }{4\pi }\int 
\mathrm{d}^{3}\mathbf{y}\hspace{0.01in}\frac{T_{\mathfrak{m}\mu \nu }(x^{0}-|%
\mathbf{x}-\mathbf{y}|,\mathbf{y})}{|\mathbf{x}-\mathbf{y}|}
\end{equation*}%
to the Einstein equations 
\begin{equation}
\square \tilde{h}_{\mu \nu }=-\kappa T_{\mathfrak{m}\mu \nu }  \label{df}
\end{equation}%
(in the same approximation) and the rest $\tilde{h}_{\mu \nu }^{\text{f}}$,
which is due to the fakeons. Combining (\ref{25}) and (\ref{df}), it is easy
to find that the difference $\tilde{h}_{\mu \nu }^{\text{f}}$ solves the
equation 
\begin{equation}
\square \left( \tilde{h}_{\mu \nu }^{\text{f}}+\frac{\kappa \eta _{\mu \nu }%
}{6m_{\chi }^{2}}\langle T_{\mathfrak{m}}\rangle _{\chi }-\frac{\kappa \eta
_{\mu \nu }}{6m_{\phi }^{2}}\langle T_{\mathfrak{m}}\rangle _{\phi }\right)
=\kappa \left( T_{\mathfrak{m}\mu \nu }-\langle T_{\mathfrak{m}\mu \nu
}\rangle _{\chi }\right) =\kappa \frac{\square }{m_{\chi }^{2}}\langle T_{%
\mathfrak{m}\mu \nu }\rangle _{\chi }.  \label{equa}
\end{equation}%
In the end, the solution reads 
\begin{equation}
\tilde{h}_{\mu \nu }^{\text{f}}=\frac{\kappa }{m_{\chi }^{2}}\langle T_{%
\mathfrak{m}\mu \nu }\rangle _{\chi }-\frac{\kappa \eta _{\mu \nu }}{%
6m_{\chi }^{2}}\langle T_{\mathfrak{m}}\rangle _{\chi }+\frac{\kappa \eta
_{\mu \nu }}{6m_{\phi }^{2}}\langle T_{\mathfrak{m}}\rangle _{\phi }.
\label{htf}
\end{equation}%
We see that $\tilde{h}_{\mu \nu }^{\text{f}}$ is a sum of fakeon averages,
which obey the properties derived in the previous section. Therefore, if the
source is slowly varying the corrections are short-range and do not affect
the radiation, which coincides with the one predicted by the Einstein
equations. If the source is not slowly varying, we also have contributions
such as those appearing in the second line of formula (\ref{omega}) with $%
\omega >m=m_{\phi },m_{\chi }$.

In the GSF\ theory, where $\phi $ is a physical particle, only $\chi _{\mu
\nu }$ is a fakeon, so the fakeon average $\langle \cdots \rangle _{\phi }$
that appears in the solution (\ref{htf}) is replaced by the convolution $%
\langle \cdots \rangle _{\phi \text{ret}}$ with the retarded Yukawa
potential. We obtain 
\begin{equation*}
\tilde{h}_{\mu \nu }^{\text{GSF}}=\tilde{h}_{\mu \nu }^{E}+\frac{\kappa }{%
m_{\chi }^{2}}\langle T_{\mathfrak{m}\mu \nu }\rangle _{\chi }-\frac{\kappa
\eta _{\mu \nu }}{6m_{\chi }^{2}}\langle T_{\mathfrak{m}}\rangle _{\chi }+%
\frac{\kappa \eta _{\mu \nu }}{6m_{\phi }^{2}}\langle T_{\mathfrak{m}%
}\rangle _{\phi \text{\hspace{0.01in}ret}}.
\end{equation*}

In the end, the gravitational waves do not amplify or propagate the
violation of microcausality, due to the large distances involved and the
damping exponential factors that appear in formulas (\ref{exa}), (\ref{appr}%
), (\ref{appr2}) and (\ref{bond}). The bounds on the masses $m_{\chi }$ and $%
m_{\phi }$ that we can obtain from this analysis are much less meaningful
than the bound\ (\ref{pote}) obtained from the Newton force.

The conclusion holds under the assumption (\ref{appro}) that the sources are
slowly varying in time, which means that their frequencies are much smaller
than the fakeon masses $m_{\chi }$ and $m_{\phi }$. As shown by the second
line of formula (\ref{comple}), frequencies of order $m_{\chi }$, $m_{\phi }$
or higher are not damped. Core-collapse supernovae are expected to generate
waves with frequencies up to about 10 kHz \cite{supernovae}. With fakeon
masses of the order of 10$^{12}$GeV, those frequencies are still too small,
but they become important if $m_{\chi }$, $m_{\phi }$ are around one MeV,
which we cannot exclude, yet.

We have mentioned that the effects of the fakeon average are a
microuncertainty on the source of the radiation. One might wonder why such
an uncertainty does not amplify, in the end. The answer is that all the
gravitational signals emitted by the averaged source propagate with the same
speed $c$, which implies that, at arbitrary distances, the microuncertainty
is just translated in spacetime along the light cones, but not amplified.

\section{Hubble constant and recovery of microcausality}

\setcounter{equation}{0}

\label{hubble}

In this section we study other situations where the potential violation of
microcausality is\ depressed rather than enhanced. In passing, the
investigation gives us the opportunity to illustrate some important aspects
of the classical limit of the fakeon prescription.

We consider a scalar field $\varphi $ (different from the scalar $\phi $
belonging to the graviton triplet) in the
Friedmann-Lemaitre-Robertson-Walker (FLRW) background. Before plunging into
the details, it is important to make a few comments on the properties of our theory on
nontrivial backgrounds. The spin-2 fakeon $\chi _{\mu \nu }$ is described by
an involved action \cite{absograv}, whose quadratic part is the
covariantized Pauli-Fierz action \cite{paulifierz}, plus nonminimal terms.
It is known that a Pauli-Fierz mass term can create pathologies \cite{patho}
on nontrivial backgrounds. For example, it may turn on ghost-like degrees of
freedom. Nevertheless, these problems do not arise in the theory of quantum
gravity we are studying. Recall that the Pauli-Fierz action of $\chi _{\mu
\nu }$ has the wrong overall sign (which is why $\chi _{\mu \nu }$ is
treated as a fakeon). This means that any degree of freedom turned on by its
mass term would have the right sign and be healthy. Actually, that degree of
freedom is already present in the theory (and under control): it is the
massive scalar $\phi $.

To see this, one must recall how the fields $\phi $ and $\chi _{\mu \nu }$
are introduced \cite{absograv}. One starts from the higher-derivative action%
\begin{equation}
S_{\text{QG}}=-\frac{M_{\text{Pl}}^{2}}{16\pi }\int \mathrm{d}^{4}x\sqrt{-g}%
\left[ R+\frac{1}{m_{\chi }^{2}}\left( R_{\mu \nu }R^{\mu \nu }-\frac{1}{3}%
R^{2}\right) -\frac{1}{6m_{\phi }^{2}}R^{2}\right]  \label{SQG}
\end{equation}%
(neglecting the cosmological term, for simplicity) and introduces auxiliary
fields to convert it into a two-derivative action. The scalar $\phi $ is
originated by the auxiliary field for $R^{2}$, while the tensor $\chi _{\mu
\nu }$ is originated by the auxiliary field for $R_{\mu \nu }R^{\mu \nu
}-R^{2}/3$. This means that the trace of $\chi _{\mu \nu }$ is not really an
independent field (it can also be seen as originated by an auxiliary field
for $R^{2}$), so it talks with $\phi $. Thus, a degree of freedom turned on
by the $\chi _{\mu \nu }$ Pauli-Fierz mass term is not independent, but can
be reabsorbed into $\phi $.

Other situations that lead to healthy massive Pauli-Fierz fields are known
in the literature, as in the de Rham-Gabadadze-Tolley model \cite{drgt} or
the compactification of five-dimensional theories \cite{five}.

We also stress that the concept of fakeon is more general than the theory of
quantum gravity it comes from. One can apply it to models of quantum gravity
that do not include massive spin-2 fields (renouncing renormalizability, as
in Einstein gravity), but also theories of matter fields (and fakeons) in
curved space. As already recalled, even the Higgs boson might be a fakeon 
\cite{fakeH}. In this sense, the results of this section contribute to the
analysis of the general properties of fakeons in connection with the issue
of microcausality.

Coming back to the problem of this section, the equation of a generic scalar 
$\varphi $ of mass $m$ interacting with an external source $J$ in a curved
background is%
\begin{equation*}
\frac{1}{\sqrt{-g}}\partial _{\mu }\left( \sqrt{-g}g^{\mu \nu }\partial
_{\nu }\varphi \right) +m^{2}\varphi =J.
\end{equation*}%
We study it under the assumption of homogeneity, $\varphi =\varphi (t)$, in
the FLRW background. The equation then reads 
\begin{equation}
\Sigma \varphi =\frac{J}{m^{2}},  \label{eq}
\end{equation}%
where $\Sigma $ denotes the operator 
\begin{equation*}
\Sigma =1+\frac{3H}{m^{2}}\frac{\mathrm{d}}{\mathrm{d}t}+\frac{1}{m^{2}}%
\frac{\mathrm{d}^{2}}{\mathrm{d}t^{2}},
\end{equation*}%
$H=\dot{a}(t)/a(t)$ is the Hubble parameter and $a(t)$ is the cosmic scale
factor. The Green function $G_{H}(t)$ is the solution of%
\begin{equation*}
\Sigma G_{H}(t)=\delta (t).
\end{equation*}%
If $\varphi $ is a fakeon, the solution is%
\begin{equation}
\varphi (t)=\int_{-\infty }^{+\infty }\mathrm{d}t^{\prime }\hspace{0.01in}%
G_{H}^{\text{f}}(t-t^{\prime })J(t^{\prime })\equiv \frac{1}{m^{2}}\langle
J\rangle _{\Sigma }(t),  \label{solv}
\end{equation}%
where the fakeon average is defined as%
\begin{equation}
\langle A\rangle _{X}\equiv \frac{1}{2}\left[ \left. \frac{1}{X}\right\vert
_{\text{rit}}+\left. \frac{1}{X}\right\vert _{\text{adv}}\right] A.
\label{fave}
\end{equation}

Let us recall that the fakeon prescription is originated perturbatively, in
momentum space (see e.g. \cite{causalityQG}). For this reason, it is
convenient to study the Fourier transforms $\tilde{G}_{H}(\omega )$, $\tilde{%
G}_{H}^{\text{f}}(\omega )$ of $G_{H}(t)$ and $G_{H}^{\text{f}}(t)$. The
retarded and advanced potentials are defined by shifting the frequency $%
\omega $ to $\omega \pm i\epsilon $ and taking the anti-Fourier transforms.

In the limit $H\rightarrow 0$ we find 
\begin{equation*}
\Sigma \rightarrow 1+\frac{1}{m^{2}}\frac{\mathrm{d}^{2}}{\mathrm{d}t^{2}},
\end{equation*}%
which is the one-dimensional version of the operator studied in section \ref%
{capmedia} and gives the fakeon Green function \cite{causalityQG} 
\begin{equation}
G_{0}^{\text{f}}(t)=\frac{m}{2}\sin (m|t|).  \label{h0}
\end{equation}%
Here the violation of microcausality is generically negligible due to the
rapidly oscillating behavior. At the cosmological level, on the other hand,
short time intervals can be important in the first moments of the universe,
so it is interesting to study the problem at nonzero $H$.

A situation that we can investigate exactly is the case of the vacuum
energy, where $H$ is constant. Since we can at most assume $H\sim $ constant
for a finite amount of time, we study the equation (\ref{eq}) in some
interval 
\begin{equation}
t_{1}\leqslant t\leqslant t_{2}.  \label{inter}
\end{equation}

We compare the cases where $\varphi $ is physical and $\varphi $ is fake.
The most general solution for the Fourier transform $\tilde{G}_{H}(\omega )$
is 
\begin{equation}
\tilde{G}_{H}(\omega )=-\frac{m^{2}}{(\omega -\omega _{+})(\omega -\omega
_{-})}+A(2\pi )\delta (\omega -\omega _{+})+B(2\pi )\delta (\omega -\omega
_{-}),  \label{mspace}
\end{equation}%
where $A$ and $B$ are arbitrary constants,%
\begin{equation*}
\omega _{\pm }=-\frac{3}{2}Hi\pm \sigma ,\qquad \sigma =\sqrt{m^{2}-\frac{9}{%
4}H^{2}},
\end{equation*}
and the \textquotedblleft complex delta function $\delta $%
\textquotedblright\ has to be understood as a series expansion in powers of $%
H/m$. Note that at $H\neq 0$ the nonvanishing imaginary parts of $\omega
_{\pm }$ make the prescriptions $\omega \rightarrow \omega \pm i\epsilon $
redundant.

If $\varphi $ is a physical field, $A$ and $B$ are determined by the initial
conditions. Instead, if $\varphi $ is a fakeon, we must set $A$ and $B$ to
zero, since the \textquotedblleft on-shell\textquotedblright\ contributions $%
\delta (\omega -\omega _{\pm })$ must be absent, by definition. Thus, the
fakeon Green function turns out to be 
\begin{equation}
G_{H}^{\text{f}}(t)=-\int \frac{\mathrm{d}\omega }{2\pi }\frac{m^{2}\mathrm{e%
}^{-i\omega t}}{(\omega -\omega _{+})(\omega -\omega _{-})}=m^{2}\mathrm{sgn}%
(H)\theta (Ht)\mathrm{e}^{-\frac{3}{2}Ht}\frac{\sin \left( t\sigma \right) }{%
\sigma },  \label{ght}
\end{equation}%
which we have written in a form that is explicit for both real and imaginary 
$\sigma $. Due to the theta function of (\ref{ght}), when $H$ is positive
only the past contributes to the fakeon solution 
\begin{equation}
\varphi (t)=\int_{-\infty }^{t}\mathrm{d}t^{\prime }\hspace{0.01in}G_{H}^{%
\text{f}}(t-t^{\prime })J(t^{\prime }),  \label{ft}
\end{equation}%
which means that the violation of microcausality disappears altogether. It
does not matter whether $m$ is small or large, since the result is exact.
If, on the other hand, $H$ is negative the opposite occurs.

The projection drops the delta-function contributions of formula (\ref%
{mspace}). Note that it is not straightforward to make the projection
directly in coordinate space, because the differential equation is only
defined in the interval (\ref{inter}). For example, we cannot discard
alleged \textquotedblleft runaway solutions\textquotedblright . Moreover,
the expressions of $\omega_{\pm }$ show that for $H>0$ the runaway
behavior concerns $t\rightarrow -\infty $, which makes no sense if the
universe has a beginning. Not to mention that the unknown differential
equation for $t<t_{1}$ could make the runaway behavior disappear.

Yet, formula (\ref{ft}) hides a subtlety: it requires knowledge of the
source $J$ for $t<t_{1}$. We might have that knowledge or not [we just know
that the differential equation is (\ref{eq}) with $H$ = constant for $%
t_{1}\leqslant t\leqslant t_{2}$]. What if we do not know $J$ in the far
past (e.g. if the universe has a beginning)?

To clarify this point, it is useful to consider the case where $\varphi $ is
a physical field, where the most general solution can be written as 
\begin{equation*}
\varphi (t)=\int_{t_{1}}^{t_{2}}\mathrm{d}t^{\prime }G_{H}^{\text{f}%
}(t-t^{\prime })J(t^{\prime })+\mathrm{e}^{-\frac{3}{2}Ht}\left[ A^{\prime
}\cos (\sigma t)+\frac{B^{\prime }}{\sigma }\sin (\sigma t)\right] ,
\end{equation*}%
where $A^{\prime }$ and $B^{\prime }$ are new constants. Here the problem of
knowing $J$ at times prior to $t_{1}$ does not show up, since this knowledge
is hidden into $A^{\prime }$ and $B^{\prime }$. However, when $\varphi $ is
a fakeon we do not have such constants and the problem remains.

What saves the day is that the damping factor and the oscillating behavior
of $G_{H}^{\text{f}}$ restrict the relevant contributions of the integral (%
\ref{ft}) to a little bit of future and a little bit of past around $t$. Let 
$\Delta t=\min (2/(3|H|),1/\sigma )$ for $m>3H/2$ and $\Delta t=\max
(1/|\omega _{+}|,1/|\omega _{-}|)$ for $m<3H/2$. If $J$ is regular and tends
to zero at infinity, the survining uncertainty 
\begin{equation*}
\delta J\equiv \int_{-\infty }^{t_{1}}\mathrm{d}t^{\prime }G_{H}^{\text{f}%
}(t-t^{\prime })J(t^{\prime })
\end{equation*}%
is small for all times $t\gtrsim t_{1}+\Delta t$. Thus, we can replace (\ref%
{ft}) with the approximate solution 
\begin{equation*}
\varphi (t)=\int_{t_{1}}^{t}\mathrm{d}t^{\prime }\hspace{0.01in}G_{H}^{\text{%
f}}(t-t^{\prime })J(t^{\prime })\text{\qquad for }t_{1}+\Delta t\lesssim
t\leqslant t_{2}.
\end{equation*}%
This result shows that we get predictivity in an interval that is slightly
smaller than (\ref{ft}).

In the end, we learn that when fakeons are present the differential
equations must be understood in a new way. In particular, we may have to
deal with uncertainties and fuzziness every time we use them.

When $H$ is approximately constant, we obtain an approximate solution by
replacing $H$ with $H(t)$. In ref. \cite{FLRW} it was shown that the
equations of the FLRW metric for the GFF theory coincide with the Friedmann
equations upon making the replacements $\rho -3p\rightarrow \langle \rho
-3p\rangle _{\Sigma }$ and $\rho +p\rightarrow \langle \rho +p\rangle
_{\Upsilon }$. The mass appearing in $\Sigma $ is $m_{\phi }$ and $\Upsilon
=\Sigma +6\dot{H}/m_{\phi }^{2}$ (at zero space curvature). Then the result (%
\ref{ft}) implies that when $H$ is constant and positive, as in the
primordial, inflationary phase of the universe, microcausality is restored
in all the equations of the GFF\ theory. In some sense, the positivity of
the Hubble constant determines the direction of time in the early universe.

After inflation, $H$ remains positive, but not constant. We do not have the
general solution $G_{H}^{\text{f}}(t)$ for a generic function $H(t)$.
Nevertheless, if $H\lesssim m_{\phi }$ we can neglect the time dependence of 
$H$ for intervals of time $\delta t$ much smaller than the Hubble time $%
t_{H}=1/H$. Indeed, the usual Friedmann equations imply $|\dot{H}|\lesssim
H^{2}$ for $p=w\rho $. If $H\lesssim m_{\phi }$, we have $\Sigma \sim
\Upsilon $ and the inequality $|\dot{H}|\lesssim H^{2}$ is also implied by
the GFF\ equations that follow from the classicization of quantum gravity.
Combining $|\dot{H}|\lesssim H^{2}$ with $|\delta t|\ll t_{H}$, we obtain $%
|\delta t|\ll H/|\dot{H}|$, which means $H(t)\sim H$ = constant. Then we can
repeat the arguments outlined above and reach similar conclusions. This
means that for amounts of time much smaller than the Hubble time (which is
comparable to the life of the universe), there is no violation of
microcausality in the classical limit.

Finally, note that the limits 
\begin{equation*}
G_{0^{\pm }}^{\text{f}}(t)\equiv \lim_{H\rightarrow 0^{\pm }}G_{H}^{\text{f}%
}(t)=m\theta (\pm t)\sin \left( m|t|\right)
\end{equation*}%
do not coincide with the $H=0$ Green function $G_{0}^{\text{f}}(t)$ of
formula (\ref{h0}). Actually, $G_{0}^{\text{f}}(t)=(G_{0^{+}}^{\text{f}%
}(t)+G_{0^{-}}^{\text{f}}(t))/2$. Basically, the resummation of the
expansion in powers of $H$ acts as a bifurcation.

\section{Conclusions}

\setcounter{equation}{0}

\label{conclusions}

The results of the investigations carried out in this paper are good news
for the consistency of the theory of quantum gravity of ref. \cite{LWgrav}
with data. At the same time, they mean that more efforts have to be spent to
identify ways to test the first departures from the predictions of Einstein
gravity.

The violation of microcausality is expressed by a fuzziness relation%
\begin{equation}
\Delta x^{2}\sim \frac{1}{m^{2}},  \label{unce}
\end{equation}%
where $m$ is the fakeon mass and $\Delta x$ is the invariant interval
between two events. The relation (\ref{unce}) means that events separated by
an interval $\Delta x$ of order $1/m$ cannot be chronologically ordered or
distinguished from each other. Because of the damping factor $\mathrm{e}%
^{-mr}$ of formulas (\ref{exa}), (\ref{appr}), (\ref{appr2}) and (\ref{bond}%
), under normal circumstances (\ref{unce}) basically means $|\Delta t|\sim
1/m$, i.e. time does not make sense below the Compton wavelength of the
fakeon. Although the relation $|\Delta t|\sim 1/m$ is not Lorentz invariant,
the apparent breakdown of Lorentz symmetry is of a spontaneous type, due to
the limitations of our experimental accuracies. The measurements of time
intervals are much less precise than those of space distances. Moreover, we
cannot change inertial frame at will. Actually, the subset of inertial
frames spanned by the macroscopic objects populating our galaxy and far
beyond is rather tiny. This makes our perception of the world quite limited,
if not biased. The idea of microcausality we inherit from it might just be a
blunder suggested by our partial insight and experimental inaccuracy.

Not to mention that in several situations, the violation of microcausality
disappears altogether, for a variety of reasons. For example, the positivity
of the Hubble constant makes the fakeon average causal and is ultimately
responsible for the arrow of time in the early universe.

\vskip12truept \noindent {\large \textbf{Acknowledgments}}

\vskip 2truept

We are grateful to M. Porrati for helpful discussions.

\renewcommand{\thesection}{A}

\section{Appendix}

\setcounter{equation}{0}

\label{appe}

It is interesting to prove the limit (\ref{delta}) directly in Minkowski
spacetime, to point out some nontrivial aspects of the fakeon Green function
and describe how the light cone contributions (\ref{-206}) disappear.
Consider 
\begin{equation*}
\int \mathrm{d}^{4}x\hspace{0.01in}G_{\text{f}}(x)J(x),
\end{equation*}%
where $J(x)$ denotes a test function. If we rescale $x\rightarrow x/m$, we
obtain $J(x/m)$ (which tends to $J(0)$ for $m\rightarrow \infty $ and can be
taken outside the integral) times an $m$-independent integral. The latter
must be computed with the help of a cutoff and a trick to properly account
for the light-cone contributions (\ref{-206}).

We switch to polar coordinates $(t,r,\theta ,\varphi )$, insert the cutoff $%
L $ on the $r$ integral (for $r$ large) and integrate the angles away. Then
we separate the integral into the sum of three contributions, to isolate the
light cones from the rest: ($i$) the integral for $|x^{2}|\leqslant \delta
^{2}$; ($ii$) the integral for $x^{2}\geqslant \delta ^{2}$ and ($iii$) the
integral for $x^{2}\leqslant -\delta ^{2}$, with $\delta $ arbitrarily
small. In ($i$) we use the approximation (\ref{-206}) and obtain 
\begin{equation}
2J(0)\int_{0}^{L}r^{2}\mathrm{d}r\int_{-\theta (r^{2}-\delta ^{2})\sqrt{%
r^{2}-\delta ^{2}}}^{\sqrt{r^{2}+\delta ^{2}}}\mathrm{d}t\hspace{0.01in}%
\delta (t^{2}-r^{2})=J(0)\frac{L^{2}}{2}.  \label{A}
\end{equation}%
In ($ii$) we get 
\begin{equation}
\frac{J(0)}{\pi }\int_{0}^{L}r^{2}\mathrm{d}r\int_{0}^{+\infty }\mathrm{d}s%
\frac{K_{1}\left( is\right) +K_{1}\left( -is\right) }{\sqrt{s^{2}+r^{2}}}%
=J(0)\left[ 1-\frac{L^{2}}{2}-\mathrm{e}^{-L}(L+1)\right] .  \label{B}
\end{equation}%
We have simplified this expression by switching to the variables $s,r$,
where $s=\sqrt{t^{2}-r^{2}}$ and noting that if we take $\epsilon
\rightarrow 0$, the integrand turns out to be regular for $s=0$. In
particular, we can let $\delta \rightarrow 0$ here, i.e. integrate $s$ from
0 to infinity. Finally, the integral ($iii$) vanishes, since $G_{\text{f}%
}(x)=0$ for $x^{2}<0$. Summing (\ref{A}) and (\ref{B}) and taking $%
L\rightarrow \infty $, we get $J(0)$, as we had to prove.

The nontrivial point is that the contributions (\ref{A}) from the light
cones (\ref{-206}) are divergent, but so are the bulk contributions (\ref{B}%
) and the total is finite.

\end{document}